\documentclass[12pt]{article}

\usepackage{amssymb,amsmath,amsthm,amsfonts,amscd}
\usepackage{subfigure}
\usepackage{graphicx}
\usepackage{color}
\usepackage[english]{babel} 
\usepackage[backend=biber,style=ieee]{biblatex}
\newcommand\be{\begin{equation}}
\newcommand\ee{\end{equation}}
\newcommand\bea{\begin{eqnarray}}
\newcommand\eea{\end{eqnarray}}

\newcommand{\fatalpha}{{\bf \alpha \kern -0.44em \alpha}}
\newcommand{\fatsigma}{{\bf \sigma \kern -0.54em \sigma}}
\newcommand{\tpchi}{{\bf \chi \kern -0.35em \chi}}
\newcommand{\llambda}{{\bf \lambda \kern -0.45em \lambda}}

	\title{\bf Geometric phase for mixed squeezed-coherent states}
	\author{ S. Mohammadi Almas
		\thanks{sanazmohammadi@uma.ac.ir} , G. Najarbashi \thanks{Najarbashi@uma.ac.ir}\\
		{\small Department of Physics, University of Mohaghegh Ardabili. P. O. Box 179, Ardabil, Iran.}\\
	}
	\pagebreak
	
\begin{document}
		\maketitle
		\newpage

\begin{abstract}
\par
In this paper, we investigate the geometric phase (GP) acquired by two-mode mixed squeezed-coherent states (SCSs) during unitary cyclic evolution, focusing on the influence of squeezing parameter and classiacal weight. We analyze the GP for three distinct mixed states characterized by different configurations of the SCSs. Our results reveal that increasing the squeezing parameters of individual modes compresses the GP contours in different patterns: linearly, hyperbolically, and elliptically, depending on the mixed state configuration. This behavior highlights the precision enhancement in squeezed states through uncertainty adjustment, aligning with theoretical predictions. 

{\bf PACS: 03.65.Vf, 03.65.Ud, 42.25.Kb}

{\bf Keywords: Geometric phase; squeezed-coherent state; mixed state}
\end{abstract}

\section{Introduction}
The concept of GP, initially explored by Pancharatnam in classical optics, stemmed from his analysis of interference patterns created by polarized light beams \cite{pancharatnam1956generalized}. Berry further developed this idea, highlighting its relevance in quantum mechanics, particularly regarding cyclic adiabatic evolution \cite{berry1984quantal}. This discovery sparked considerable interest in holonomy effects within quantum mechanics, leading to numerous generalizations of GP. Aharonov and Anandan \cite{aharonov1987phase} extended this concept to nonadiabatic cyclic evolution, while Samuel and Bhandari \cite{samuel1988general} broadened it by applying it to noncyclic evolution and sequential projection measurements. GP is a consequence of quantum kinematics and is therefore independent of the specific nature of the path's dynamics in state space. Mukunda and Simon \cite{mukunda1993quantum} proposed a kinematic approach, emphasizing the path traversed in state space as the primary concept for understanding GP. Subsequent generalizations and refinements relaxed the conditions of adiabaticity, unitarity, and cyclicity of the evolution \cite{pati1995geometric}. 
\par
The extension of GP to mixed states was first addressed by Uhlmann \cite{uhlmann1986parallel} within a mathematical framework of purification. Sjöqvist et al. \cite{sjoqvist2000interferometry} proposed an alternative definition based on quantum interferometry, while Singh et al. \cite{singh2003geometric} offered a kinematic description of mixed state GP and extended it to degenerate density operators. Additionally, the relationship between phases of entangled systems and their subsystems has been explored \cite{sjoqvist2000geometric, tong2003relation, tong2003geometric, ericsson2003mixed}.
\par
Another line of development has been towards extending the GP to coherent and squeezed states. Coherent states, introduced by Glauber in 1963 \cite{glauber1963coherent}, laid a crucial foundation for investigating the quantum properties of light. In 1976, Yuen introduced squeezed states, which are characterized by reduced quantum noise in one quadrature of the electromagnetic field compared to the vacuum state \cite{yuen1976two}. The initial study of the GP in coherent states was conducted by Klauder et al. \cite{klauder1985coherent, klauder2006fundamentals}. Building on this, Kuratsuji et al. \cite{kuratsuji1985effective} explored the adiabatic GP in coherent states, with Kuratsuji and Littlejohn\cite{kuratsuji1988geometric, littlejohn1988cyclic} later extending this work to non-adiabatic evolutions. This line of research has led to extensive studies of non-cyclic GPs in both coherent and squeezed states \cite{mendas1997pancharatnam, chaturvedi1987berry, pati1995geometric, sjoqvist1997noncyclic, yang2011geometric}. Additionally, the GP of entangled coherent and squeezed states has been investigated. The GP of two-mode entangled coherent states has been analyzed in relation to their concurrence \cite{Almas2022geometric}. It was found that balanced and unbalanced entangled coherent states with the same degree of entanglement can exhibit different GPs, suggesting that coherent states with identical entanglement levels can have varying GPs. The influence of the squeezing parameter on the GP for two-mode entangled SCSs has been a topic of investigation as well, highlighting the interplay between squeezing and the GP in quantum systems \cite{mohammadi2024geometric}. 
\par
While earlier research focused on pure SCSs \cite{mohammadi2024geometric}, in this paper we extend the analysis to mixed SCSs. This generalization allows us to explore whether the previously observed results for pure states hold in the mixed state context, thus providing deeper insights into the robustness of the relationship between squeezing and GP in more complex quantum systems.
\par
In this paper, we focus on the GP of mixed SCSs subjected to unitary cyclic evolution. SCSs are particularly important in enhancing the security of quantum cryptographic protocols by providing improved protection against eavesdropping and increasing the efficiency of information transmission. For instance, the study by Jeong et al. \cite{jeong2015gaussian} discusses how SCSs can improve the security of Gaussian private quantum channels, thereby enhancing security in quantum key distribution. Additionally, squeezed states have been investigated for their capability to enhance phase estimation precision in lossy quantum optical metrology, demonstrating significant improvements in measurement accuracy even in the presence of substantial photon loss \cite{zhang2013lossy}.
\par
GPs play a significant role in quantum information processing and quantum computation \cite{zanardi1999holonomic, jones2000geometric, zhu2002implementation, vedral2003geometric, rowell2018mathematics}. Recently, it has been shown that the use of GP in a clock interferometer can enhance precision in metrology \cite{zhou2024geometric}, where leveraging GP has led to significant improvements in the accuracy of metrological measurements. Choi's work \cite{choi2020quadrature} further explores the time evolution of GPs in squeezed light states within nano-optics, demonstrating that squeezing induces novel GP oscillations. These oscillations, which can be finely controlled by adjusting the squeezing parameters, also reveal additional behaviors in one-photon light-matter interactions. Despite the importance of GPs, their applications in SCSs have been discussed less frequently in the literature. Studying the GP in SCSs can lead to the development of new techniques and protocols for implementing quantum gates and operations that exploit the unique properties of these states. Furthermore, the manipulation and control of the GP for SCSs offer avenues to enhance quantum measurement precision, thus paving the way for advancements in high-precision sensing, quantum interferometry, and a wide range of metrological applications.
\par
The organization of this paper is as follows: Section 2 provides a brief overview of the kinematic approach underpinning the GP for mixed states in the context of unitary evolutions. The foundational concepts and mathematical formulations are introduced, setting the stage for subsequent discussions. Section 3 delves into the exploration of GP attributes within two-mode mixed SCSs during unitary cyclic evolution, further dissecting the role of squeezing parameters. Section 4 extends analysis of the results obtained in the previous sections. The paper ends with a summary and conclusions in Section 5.

\section{Quantum kinematic approach to the GP for mixed states}
Let's consider a mixed state density matrix $\rho(0)$ that evolves along an open unitary path $\mathcal{C}$ from $t$ to $\tau$, where $\rho(t) = U(t)\rho(0)U^\dagger(t)$ in the space of density operators, starting at $\rho(0)	$ and ending at $\rho(\tau)$. This unitary evolution is not necessarily cyclic, meaning $\rho(\tau) \neq \rho(0)$. The total phase acquired by the mixed state during this evolution, as detailed in \cite{sjoqvist2000interferometry}, is given by:
\begin{equation}\label{total}
\gamma_{T}= \arg \mathrm{Tr}[\rho(0) U(t)].
\end{equation}
The dynamical phase, can be defined as \cite{sjoqvist2000interferometry}:
\begin{equation}\label{dyn}
\gamma_{D}=-i\int\limits_{0}^{\tau} dt \mathrm{Tr}[\rho(0) U^\dagger(t) \dot{U}(t)].
\end{equation}
The GP associated with the projective Hilbert space is then the difference between the total accumulated phase and the dynamical phase, expressed as \cite{mukunda1993quantum}:
\begin{equation}\label{gp}
\gamma_{G}[\mathcal{C}]=\arg \mathrm{Tr}[\rho(0) U(t)]+i\int\limits_{0}^{\tau} dt \mathrm{Tr}[\rho(0) U^\dagger(t) \dot{U}(t)].
\end{equation}
The GP is a unique feature associated with the path $\mathcal{C}$ taken through the projective Hilbert space. It remains unchanged when subjected to gauge transformations or reparametrizations \cite{mukunda1993quantum}.

\section{GP of mixed SCSs}
\par
In this section, we explore the GP acquired by two-mode mixed SCSs during unitary evolution. The primary aim is to investigate how these states behave under cyclic evolution and to compare the resulting GPs. To begin, we briefly outline the theoretical framework that underpins the construction of SCSs.
\par
SCSs are generated by applying two key quantum operators to the vacuum state: the squeezing operator, $\hat{S}(\xi)$, and the displacement operator, $\hat{D}(\alpha)$. These operators, when applied consecutively, modify the vacuum state to produce a SCS, expressed as:
\begin{equation}
|\alpha,\xi\rangle=\hat{D}(\alpha)\hat{S}(\xi)|0\rangle.
\end{equation}
The squeezing operator is represented by the expression $\exp[\frac{1}{2}(\xi^{\ast} \hat{a}^{2} - \xi \hat{a}^{\dagger 2})]$, where $\xi = r e^{i \Theta}$ defines the squeezing parameter $r$ and the squeezing angle $\Theta$. The squeezing parameter $r$ controls the degree of quantum noise reduction in one quadrature at the expense of increased noise in the conjugate quadrature. On the other hand, the displacement operator is written as $\exp(\alpha \hat{a}^{\dagger} - \alpha \hat{a})$, where $\alpha$ is a complex number referred to as the coherence parameter, determining the displacement of the state from the origin in phase space.
\par
If $\alpha = 0$, the state reduces to a purely squeezed state. Conversely, when $\xi = 0$, the state becomes a regular coherent state. The combination of these two operators produces the general SCS, $|\alpha, \xi\rangle$, which exhibits both displacement and squeezing effects.
\par
Moreover, SCSs can be further characterized by their eigenvalue equation with respect to the operator $\hat{A} = \hat{a} \cosh r + \hat{a}^{\dagger} e^{i \Theta} \sinh r$, which leads to the following eigenstate relationship \cite{gerry2005introductory}:
\begin{equation}\label{eigenstate-equation}
\hat{A}|\alpha,\xi\rangle=\eta|\alpha,\xi\rangle,
\end{equation}
where $\eta = \alpha \cosh r + \alpha e^{i \Theta} \sinh r$. In the special case where both $\alpha$ and $\xi$ are real, this simplifies to $\eta = \alpha e^{r}$, highlighting the direct relationship between the squeezing parameter and the state’s displacement in phase space.
\par
An important representation of an SCS involves expressing it in the Fock (number state) basis $|n\rangle$. The state $|\alpha, \xi\rangle$ can be expanded as \cite{gerry2005introductory}:
\begin{align}
|\alpha,\xi\rangle=&\frac{1}{\sqrt{\cosh r}}\exp[-\frac{1}{2}|\alpha|^{2}-\frac{1}{2}\alpha^{\ast2}e^{i\Theta}\tanh r] \\ \nonumber
&\times \sum\limits_{n=0}^\infty \frac{(\frac{1}{2}e^{i\Theta}\tanh r)^{n/2}}{\sqrt{n!}}H_{n}[\eta(e^{i\Theta}\sinh(2r))^{-1/2}]|n\rangle,
\end{align}
where $H_n$ represents the Hermite polynomials, and the sum runs over all possible Fock states. This expression shows how the state is a superposition of number states, weighted by the Hermite polynomials, and is influenced by the squeezing and displacement parameters.
\par
The overlap between two SCSs, $|\alpha_0, \xi_0\rangle$ and $|\alpha_1, \xi_1\rangle$, is given by the following expression, assuming the parameters $\alpha$ and $\xi$ are real:
\begin{align}
\langle\alpha_{0},\xi_{0}|\alpha_{1},\xi_{1}\rangle=&\frac{1}{\sqrt{\cosh r_{0}\cosh r_{1}}}\exp[-\frac{1}{2}\alpha_{0}^{2}(1+\tanh r_{0})-\frac{1}{2}\alpha_{1}^{2}(1+\tanh r_{1})] \\ \nonumber
&\times\sum\limits_{n=0}^{\infty}\frac{1}{2^{n}n!} (\tanh r_{0}\tanh r_{1})^{n/2} H_{n}[\alpha_{0}e^{r_{0}}(\sinh(2r_{0}))^{-1/2}]H_{n}[\alpha_{1}e^{r_{1}}(\sinh(2r_{1}))^{-1/2}].
\end{align}
To simplify the summation over the Hermite polynomials, we apply Mehler's formula \cite{beals2016special}, a well-known result for sums of products of Hermite polynomials:
\begin{align}
\sum\limits_{n=0}^{\infty}\frac{H_{n}[x]H_{n}[y]s^{n}}{2^{n}n!}=\frac{1}{\sqrt{1-s^{2}}}\exp[\frac{2xys-x^{2}s^{2}-y^{2}s^{2}}{1-s^{2}}],
\end{align}
where $|s| < 1$. This identity significantly reduces the complexity of the expression for the overlap between two SCSs. Using this formula, the inner product between the two SCSs simplifies to:
\begin{align}\label{scalarproduct}
\langle\alpha_{0},\xi_{0}|\alpha_{1},\xi_{1}\rangle=&\frac{1}{\sqrt{\cosh (r_{0}-r_{1})}}\exp[-\frac{1}{2}\alpha_{0}^{2}(1+\tanh r_{0})-\frac{1}{2}\alpha_{1}^{2}(1+\tanh r_{1})] \\ \nonumber
&\times\exp[\frac{\alpha_{0}\alpha_{1}e^{(r_{0}+r_{1})}}{\cosh (r_{0}-r_{1})}-\frac{\alpha_{0}^{2}e^{2r_{0}}\sinh r_{1}}{2\cosh r_{0}\cosh (r_{0}-r_{1})}-\frac{\alpha_{1}^{2}e^{2r_{1}}\sinh r_{0}}{2\cosh r_{1}\cosh (r_{0}-r_{1})}].
\end{align}
This result gives the overlap in terms of both the squeezing parameters and the displacement parameters. The conditions $0 < \tanh r_0 \tanh r_1 < 1$ ensure that the squeezing effects remain within a physically valid regime.

\subsection{Entangled balanced-unbalanced mixed SCS}
In this subsection, we delve into the examination of the GP acquired by an two-mode mixed SCS, emphasizing its behavior during unitary cyclic evolution. Our focus is on understanding how the GP manifests in these states as various parameters change.
\par
We start by considering an initial mixed state that is a composition of two distinct SCSs $|\psi_{1}\rangle$ and $|\psi_{2}\rangle$. These states are distinct pure SCSs within the two-mode system. The mixed state can be represented by the density matrix:
\begin{equation}\label{ESCS}
\rho(0)=\lambda |\psi_{1}\rangle \langle \psi_{1}|+(1-\lambda) |\psi_{2}\rangle \langle \psi_{2}|,
\end{equation}
where $0 \leq \lambda \leq 1$ is the classical weight that determines the degree of mixing between the two states $|\psi_{1}\rangle$ and $|\psi_{2}\rangle$. The parameter $\lambda$ effectively tunes the contribution of each state to the mixed state $\rho(0)$.
\par
For this analysis, we assume the states $|\psi_{1}\rangle$ and $|\psi_{2}\rangle$ are defined as follows:
\begin{align}
\begin{array}{c}
|\psi_{1}\rangle=\frac{1}{\sqrt{N}}(|\alpha_{0},\xi_{0}\rangle|\alpha_{0},\xi_{0}\rangle+|\alpha_{1},\xi_{1}\rangle|\alpha_{1},\xi_{1}\rangle),\\ 
|\psi_{2}\rangle=\frac{1}{\sqrt{N}}(|\alpha_{0},\xi_{0}\rangle|\alpha_{1},\xi_{1}\rangle+|\alpha_{1},\xi_{1}\rangle|\alpha_{0},\xi_{0}\rangle), \\
\end{array}
\end{align}
where $|\psi_{1}\rangle$ represents a balanced entangled SCS, with the terms being symmetrically entangled, and $|\psi_{2}\rangle$ represents an unbalanced entangled SCS, with the terms being asymmetrically entangled. The normalization factor $N$ is given by $N=2+2p_{01}$, where $p_{01}=\langle\alpha_{0},\xi_{0}|\alpha_{1},\xi_{1}\rangle$ represents the overlap between the two SCSs. For simplicity, we consider all parameters to be real numbers.

\par
Next, we aim to compute the GP when the introdued mixed state undergoes unitary cyclic evolution. The operator $\hat{A}$, as defined in Eq. (\ref{eigenstate-equation}), follows the commutation relation $[\hat{A}, \hat{A}^\dagger] = 1$. By utilizing the Jordan-Schwinger representation of the $SU(2)$ algebra, we introduce two sets of bosonic operators ${\hat{A}, \hat{A}^\dagger}$ and ${\hat{B}, \hat{B}^\dagger}$, corresponding to the first and second modes of the system. These operators satisfy $[\hat{A}, \hat{A}^\dagger] = 1$, $[\hat{B}, \hat{B}^\dagger] = 1$, and commute with each other, i.e., $[\hat{A}, \hat{B}] = 0$.
\par
With these bosonic operators, we define three Hermitian operators $\hat{J}_x$, $\hat{J}_y$, and $\hat{J}_z$, which correspond to the components of angular momentum:
\begin{align}
\begin{array}{c}
{\hat{J_{x}}=\frac{1}{2}(\hat{A}^{\dag}\hat{B}+\hat{A}\hat{B}^{\dag})},\\ {\hat{J_{y}}=\frac{1}{2i}(\hat{A}^{\dag}\hat{B}-\hat{A}\hat{B}^{\dag})},\\ {\hat{J_{z}}=\frac{1}{2}(\hat{A}^{\dag}\hat{A}-\hat{B}^{\dag}\hat{B})}. \\
\end{array}
\end{align}
These operators satisfy the familiar angular momentum commutation relations: $[\hat{J}_i, \hat{J}j] = i \varepsilon_{ijk} \hat{J}_{k}$, where $\varepsilon_{ijk}$ is the Levi-Civita symbol and $i, j, k$ represent different Cartesian components. We work in units where $\hbar = 1$. The vector $\hat{J} = ({\hat{J_{x}},\hat{J_{y}},\hat{J_{z}}})$ describes the total angular momentum of the two-mode system.
\par
To evolve the state unitarily, we apply a rotation operator of the form $e^{-i \phi \hat{J} \cdot \hat{n}}$, which performs a rotation around the axis $\hat{n}$ by an angle $\phi$. In particular, we use a product of two rotation operators to define the unitary evolution:
\begin{equation}\label{unitaryevolution}
\hat{U} (\theta,\varphi)=e^{-i\varphi\hat{J_{z}}}e^{-i\theta\hat{J_{y}}},
\end{equation}
which corresponds to a rotation around the $y$-axis by angle $\theta$, followed by a rotation around the $z$-axis by angle $\varphi$. This operator encapsulates a non-local unitary transformation of the two-mode system.
\par
Next, we consider the cyclic evolution of the system by fixing $\theta$ and allowing $\varphi$ to vary from $0$ to $2\pi$. Under this evolution, the trace $\mathrm{Tr}[\rho(0) U(t)]$ remains real and positive, leading to the total phase $\gamma_T$ being zero. To compute the dynamical phase $\gamma_D$, we calculate the quantity $\hat{U}^\dagger(\theta, \varphi) \partial_\varphi \hat{U}(\theta, \varphi)$. Using the Baker-Campbell-Hausdorff formula, we find:
\begin{equation}\label{unitary}
\hat{U}^{\dag}(\theta,\varphi)\partial_{\varphi}\hat{U}(\theta,\varphi)=-i (\cos\theta \hat{J_{z}}-\sin\theta \hat{J_{x}}).
\end{equation}
Substituting this into the expression for the dynamical phase (as given by Eq. (\ref{dyn})), and integrating over $\varphi$ from $0$ to $2\pi$, we obtain the following expression for the dynamical phase $\gamma_D$ for a fixed $\theta$:
\begin{equation}
\gamma_{D}(\rho)=\frac{2\pi \sin\theta}{N}\{\lambda(\eta_{0}^{2}+\eta_{1}^{2}+2\eta_{0}\eta_{1}p_{01}^{2})+(1-\lambda)((\eta_{0}^{2}+\eta_{1}^{2})p_{01}^{2}+2\eta_{0}\eta_{1})\},
\end{equation}
where the parameters $\eta_0 = \alpha_0 e^{r_0}$ and $\eta_1 = \alpha_1 e^{r_1}$ are related to the SCSs $|\alpha_0, \xi_0\rangle$ and $|\alpha_1, \xi_1\rangle$, respectively. 
\par
Finally, after obtaining the total phase and dynamical phase, we use Eq.(\ref{gp}) to find the GP of the miexd state. The GP is given by: 
\begin{equation}\label{gp-mixstate1}
\gamma_{G}(\rho)=\frac{-2\pi \sin\theta}{N}\{\lambda(\eta_{0}^{2}+\eta_{1}^{2}+2\eta_{0}\eta_{1}p_{01}^{2})+(1-\lambda)((\eta_{0}^{2}+\eta_{1}^{2})p_{01}^{2}+2\eta_{0}\eta_{1})\}.
\end{equation}
This expression shows that the GP depends on the angle $\theta$, the classical weight $\lambda$, and the coherence and squeezing parameters associated with the two modes of the initial mixed state.
\par
The parameter $\theta$ represents the angle in the unitary rotation operator. The dependency on $\theta$ suggests that different paths in the parameter space lead to different GPs. The classical weight $\lambda$ controls the contribution of each SCS to the mixed state. When$\lambda=0$, the state is the pure state $|\psi_{2}\rangle$, and when $\lambda=1$, it is the pure state $|\psi_{1}\rangle$. The GP smoothly interpolates between these two extreme cases, highlighting the dependency on the classical weight.
\par
\begin{figure}
 \centering
 \subfigure[$\lambda=0$]{\includegraphics[width=5cm]{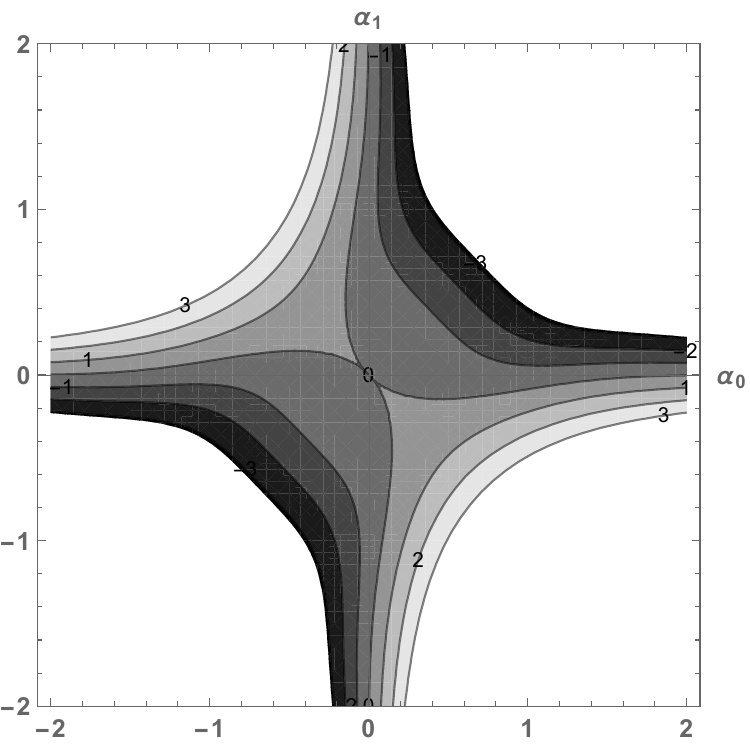}\label{1a}}
  \hfill
 \subfigure[$\lambda=1/4$]{\includegraphics[width=5cm]{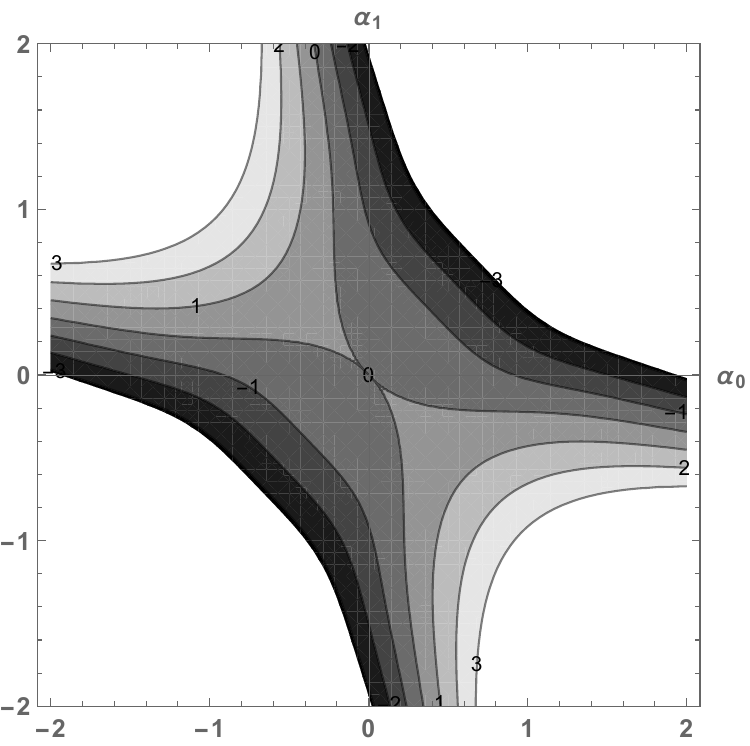}\label{1b}}
  \hfill
 \subfigure[$\lambda=1/2$]{\includegraphics[width=5cm]{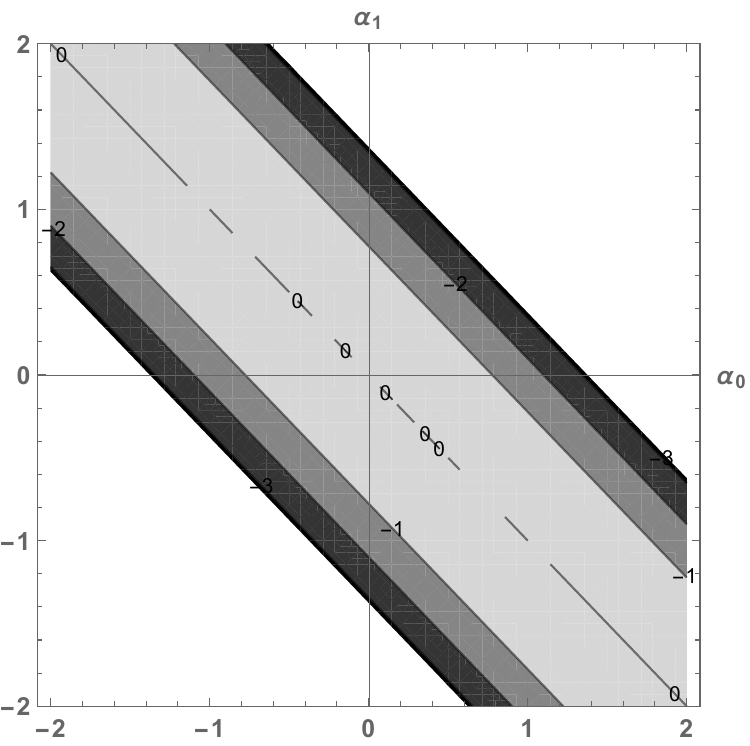}\label{1c}}
  \hfill
 \subfigure[$\lambda=3/4$]{\includegraphics[width=5cm]{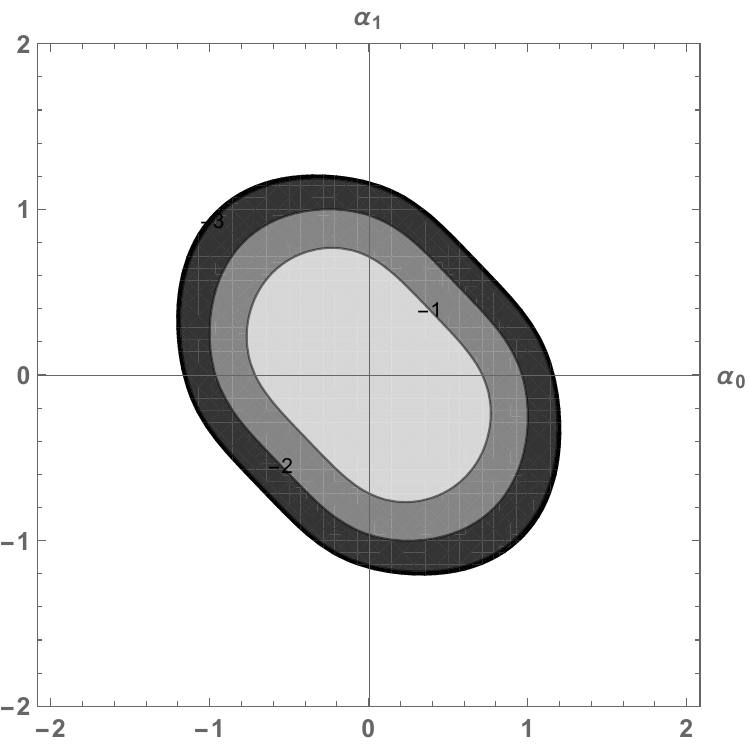}\label{1d}}
\hspace{1.5 em}
  \subfigure[$\lambda=1$]{\includegraphics[width=5cm]{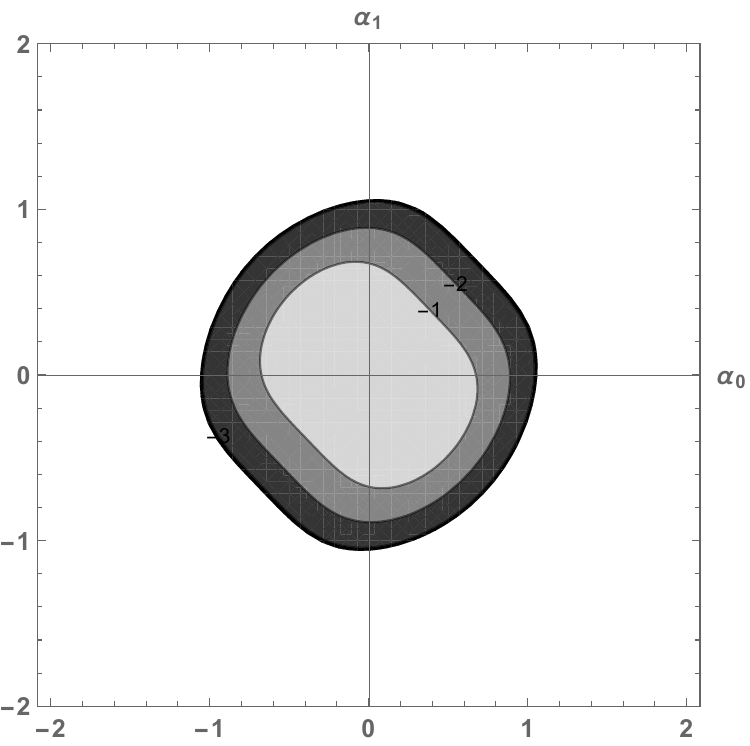}\label{1e}}

  \caption{\small{Contour plots of the GP for $\rho$ as functions of $\alpha_{0}$ and $\alpha_{1}$, for $\theta=\pi/4$, $r_{0}=r_{1}=0.2$ and various values of the classical weight.}}\label{ContourgpSCS1}
 \end{figure}

When $\lambda=0$, the GP in Eq.(\ref{gp-mixstate1}) corresponds to that of state $|\psi_{2}\rangle$, whereas for $\lambda=1$, it corresponds to $|\psi_{1}\rangle$. The GP for pure states of $|\psi_{1}\rangle$ and $|\psi_{2}\rangle$ has been studied in \cite{mohammadi2024geometric}. To visually illustrate the GP's behavior, Fig. \ref{ContourgpSCS1} presents contour plots depicting the GP for various values of the classical weight, while keeping $\theta=\pi/4$ and $r_{0}=r_{1}=0.2$. The figures show that for $\lambda=0$, the plot exhibits a hyperbolic-like shape, whereas for $\lambda=1$, it displays an elliptic-like shape. As $\lambda$ increases continuously from $0$ to $1$, the plot transitions from a hyperbolic-like to an elliptic-like form. This transition demonstrates the ability to manipulate and control the GP through the classical weight in practical applications involving mixed SCSs.

\subsection{Separable unbalanced mixed SCS}
Now, we consider another two-mode mixed SCS, where the state introduced in Eq. (\ref{ESCS}) is characterized by different configurations of the SCSs $|\psi_{1}\rangle$ and $|\psi_{2}\rangle$. Specifically, we assume the states are $|\psi_{1}\rangle=|\alpha_{0},\xi_{0}\rangle|\alpha_{1},\xi_{1}\rangle$ and $|\psi_{2}\rangle=|\alpha_{1},\xi_{1}\rangle|\alpha_{0},\xi_{0}\rangle$, both of which represent separable unbalanced SCSs. We denote this initial mixed state as $\rho^{\prime}(0)$, which can be expressed as:
\begin{equation}\label{unbalSCS}
\rho^{\prime}(0)=\lambda (|\alpha_{0},\xi_{0}\rangle|\alpha_{1},\xi_{1}\rangle \langle \alpha_{0},\xi_{0}|\alpha_{1},\xi_{1}|)+(1-\lambda) (|\alpha_{1},\xi_{1}\rangle|\alpha_{0},\xi_{0}\rangle \langle \alpha_{1},\xi_{1}|\alpha_{0},\xi_{0}|).
\end{equation}
\par
\begin{figure}
 \centering
 \subfigure[$\lambda=0$]{\includegraphics[width=5cm]{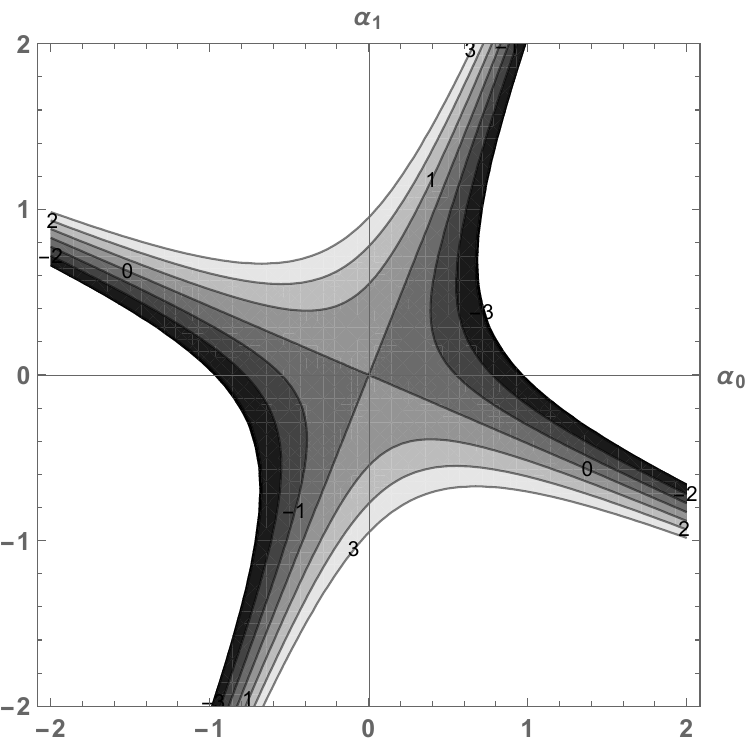}\label{2a}}
  \hfill
 \subfigure[$\lambda=1/4$]{\includegraphics[width=5cm]{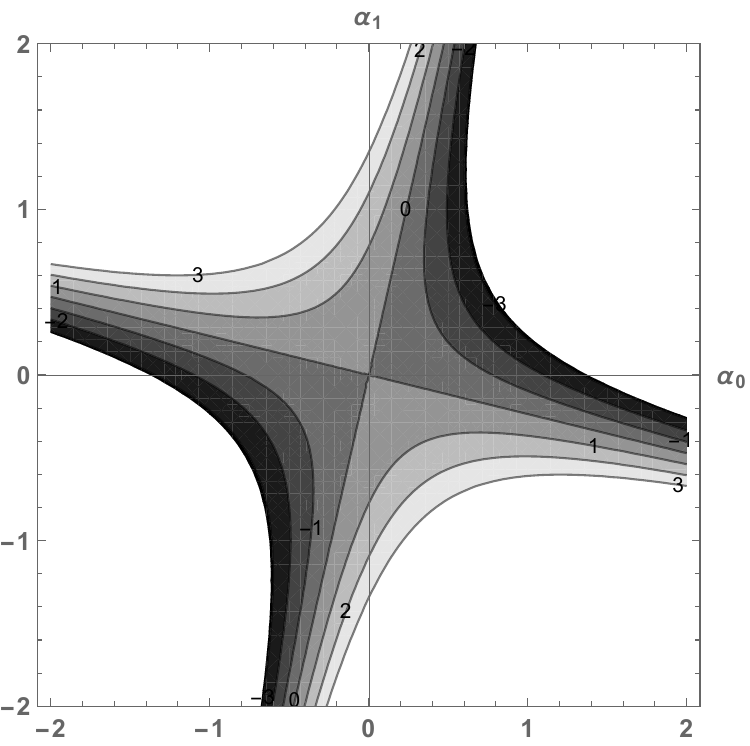}\label{2b}}
  \hfill
 \subfigure[$\lambda=1/2$]{\includegraphics[width=5cm]{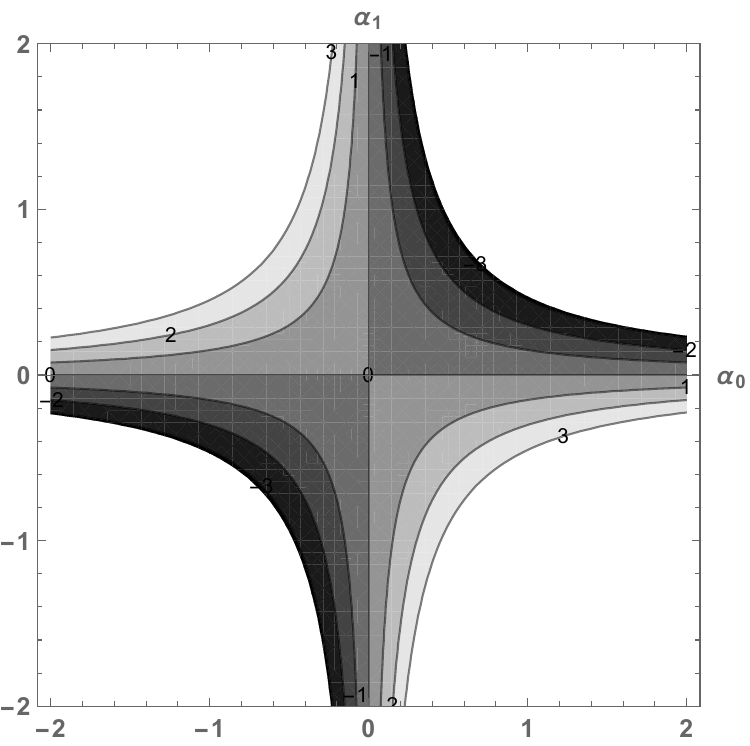}\label{2c}}
  \hfill
 \subfigure[$\lambda=3/4$]{\includegraphics[width=5cm]{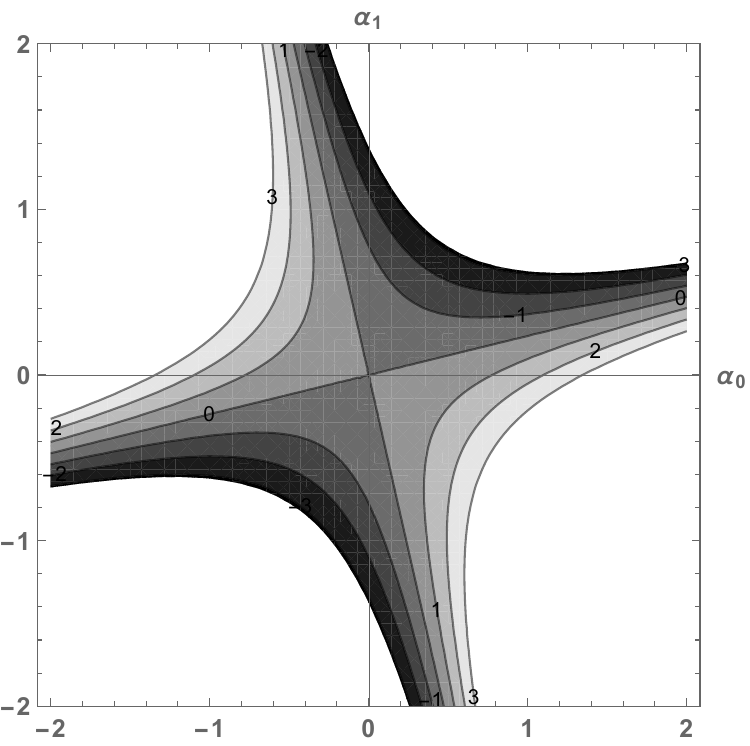}\label{2d}}
 \hspace{1.5 em}
  \subfigure[$\lambda=1$]{\includegraphics[width=5cm]{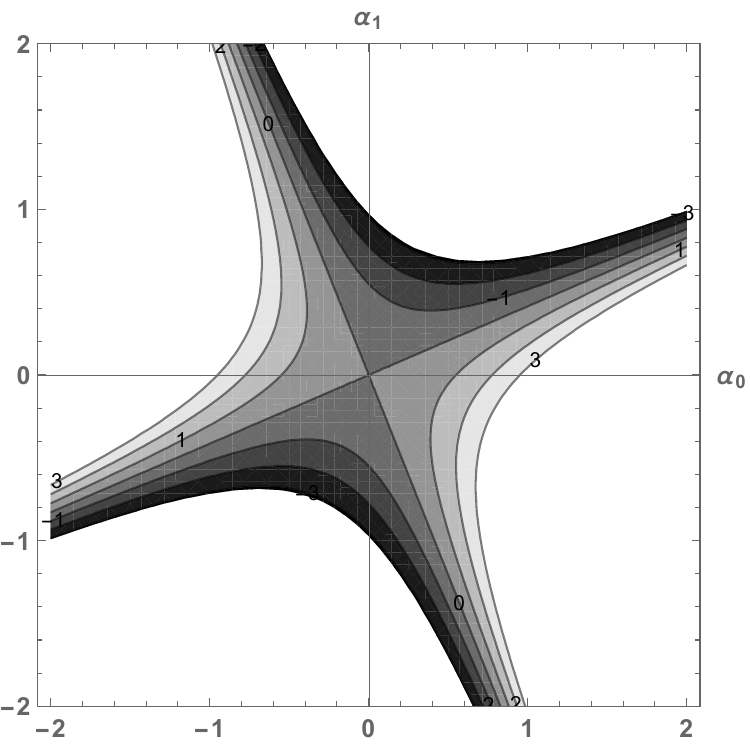}\label{2e}}

\caption{\small{Contour plots of the GP for $\rho^{\prime}$ as functions of $\alpha_{0}$ and $\alpha_{1}$, for $\theta=\pi/4$, $r_{0}=r_{1}=0.2$ and various values of the classical weight.}}\label{ContourgpSCS2}
 \end{figure}

To understand the geometric properties of this mixed state, we calculate the GP it acquires during a cyclic unitary evolution described by Eq. (\ref{unitaryevolution}). As in the previous subsection, it can be checked that the total phase of the system is equal to zero. By calculating the dynamical phase, we can then determine the GP. Following similar steps as before, we find that the GP for the mixed state $\rho^{\prime}$ is given by:
\begin{equation}\label{gp-mixstate2}
\gamma_{G}(\rho^{\prime})=2\pi\{-\eta_{0}\eta_{1} \sin\theta+(\eta_{0}^{2}-\eta_{1}^{2})(\lambda-1/2)\cos\theta\}.
\end{equation}
When $\lambda=0$, the state is the pure state $|\psi_{2}\rangle$, and when $\lambda=1$, it is the pure state $|\psi_{1}\rangle$. The GP smoothly interpolates between these two cases, emphasizing the dependency on the degree of mixing between the two separable SCSs. This interpolation illustrates how varying the classical weight $\lambda$ allows for a continuous transition in the GP, reflecting the gradual change in the state's composition from one pure state to the other (See the Fig. \ref{ContourgpSCS2}).

\subsection{Separable balanced mixed SCS}
We consider a different configuration of a two-mode mixed SCS. In this case, the mixed state introduced in Eq. (\ref{ESCS}) is characterized by separable balanced SCSs, specifically $|\psi_{1}\rangle=|\alpha_{0},\xi_{0}\rangle|\alpha_{0},\xi_{0}\rangle$ and $|\psi_{2}\rangle=|\alpha_{1},\xi_{1}\rangle|\alpha_{1},\xi_{1}\rangle$. We represent this initial mixed state as $\rho^{\prime \prime}(0)$, which can be written as:
\begin{equation}\label{balSCS}
\rho^{\prime \prime}(0)=\lambda (|\alpha_{0},\xi_{0}\rangle|\alpha_{0},\xi_{0}\rangle \langle \alpha_{0},\xi_{0}|\alpha_{0},\xi_{0}|)+(1-\lambda) (|\alpha_{1},\xi_{1}\rangle|\alpha_{1},\xi_{1}\rangle \langle \alpha_{1},\xi_{1}|\alpha_{1},\xi_{1}|).
\end{equation}
We assume that this state is also subjected to the unitary evolution described by Eq. (\ref{unitaryevolution}). Considering a fixed $\theta$ and cyclic evolution of $\varphi$ from $0$ to $2\pi$, the GP is given by: 
\begin{equation}\label{gp-mixstate3}
\gamma_{G}(\rho^{\prime \prime})=-2\pi \sin\theta \{\lambda \eta_{0}^{2}+(1-\lambda)\eta_{1}^{2}\}.
\end{equation}
Like the previous two states, the total phase of this state also becomes zero in a closed path, implying that $\gamma_{G}=-\gamma_{D}$. The GP is influenced by the degree of mixing between the two balanced SCSs, the angle $\theta$ and the coefficients $\eta_{0}$ and $\eta_{1}$, which are functions of the coherence and squeezing parameters of the individual modes.
\par
\begin{figure}
 \centering
 \subfigure[$\lambda=0$]{\includegraphics[width=5cm]{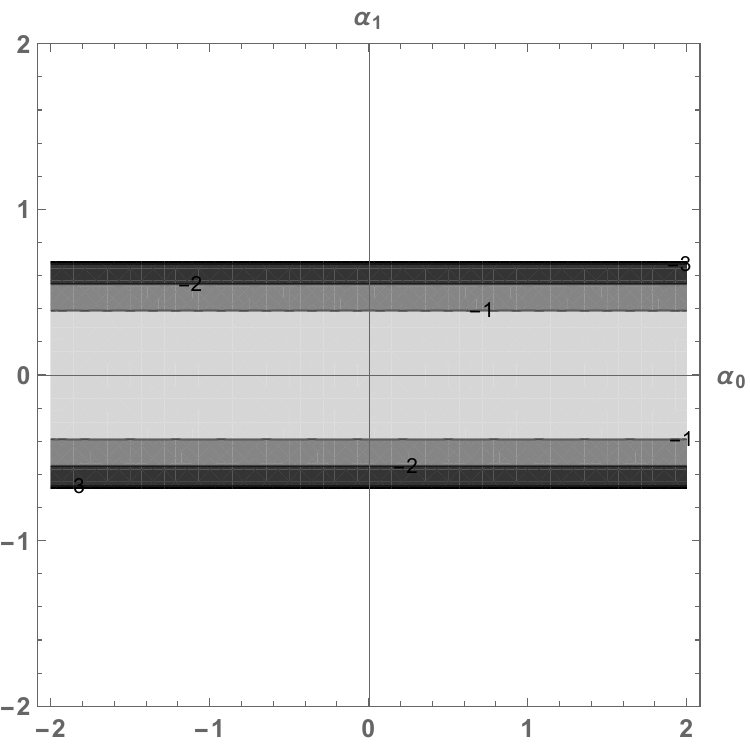}\label{3a}}
  \hfill
 \subfigure[$\lambda=1/4$]{\includegraphics[width=5cm]{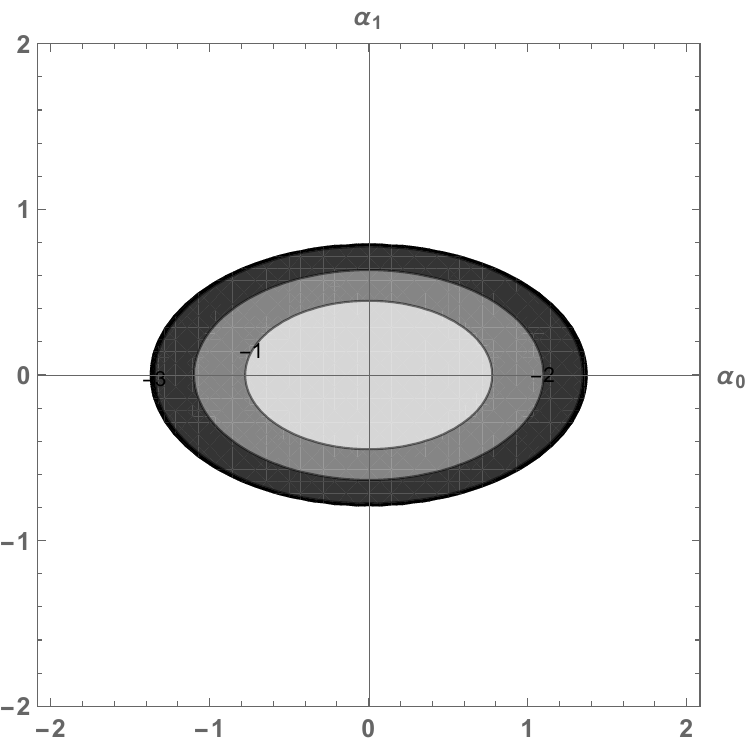}\label{3b}}
\hfill
 \subfigure[$\lambda=1/2$]{\includegraphics[width=5cm]{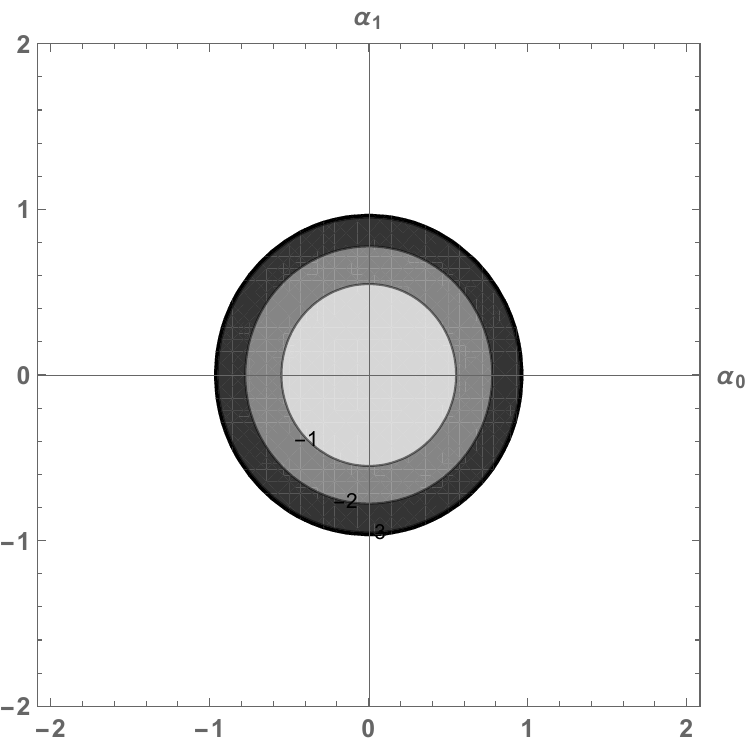}\label{3c}}
 \hfill
 \subfigure[$\lambda=3/4$]{\includegraphics[width=5cm]{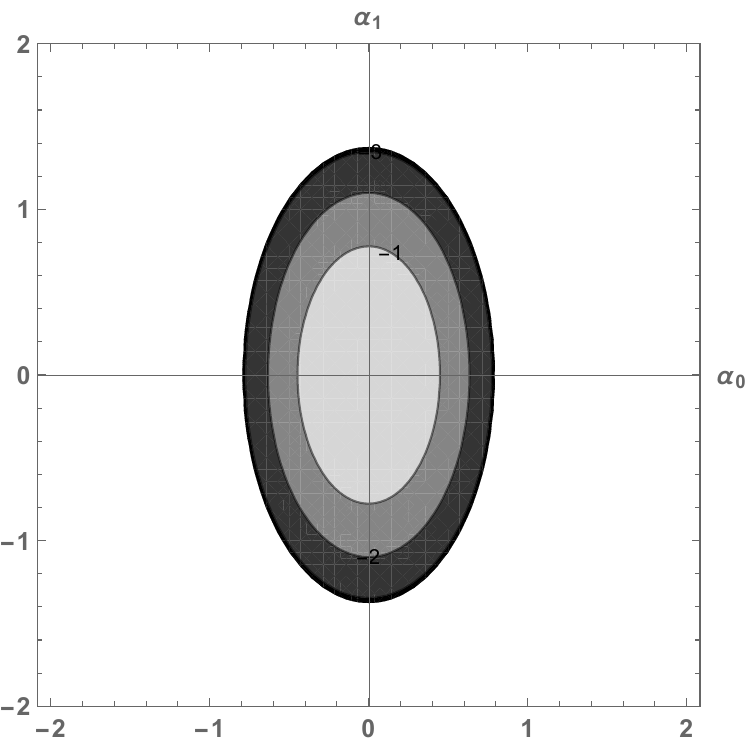}\label{3d}}
\hspace{1.5 em}
  \subfigure[$\lambda=1$]{\includegraphics[width=5cm]{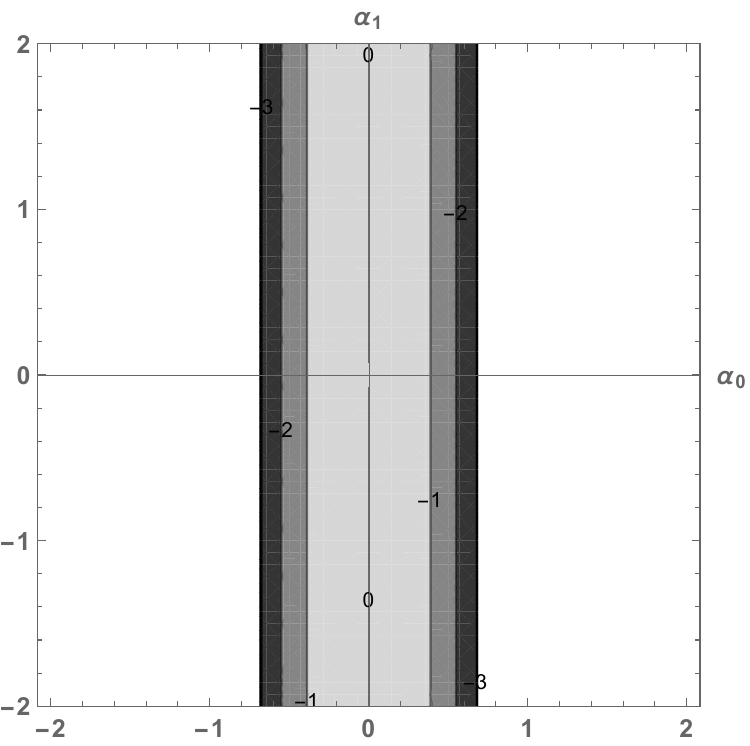}\label{3e}}

  \caption{\small{Contour plots of the GP for $\rho^{\prime\prime}$ as functions of $\alpha_{0}$ and $\alpha_{1}$, for $\theta=\pi/4$, $r_{0}=r_{1}=0.2$ and various values of the classical weight.}}\label{ContourgpSCS3}
 \end{figure}

In Fig. \ref{ContourgpSCS3}, the calculated GPs for different values of the classical weight $\lambda$ are plotted as functions of $\alpha_{0}$ and $\alpha_{1}$ for the mixed state $\rho^{\prime\prime}$. When $\lambda=0$, the plot exhibits a linear-like shape, indicating that the GP is primarily influenced by the state $|\psi_{2}\rangle$. As $\lambda$ increases continuously from $0$ to $1$, the plot transitions through various shapes, including hyperbolic-like contours. At $\lambda=1$, the plot returns to a linear-like shape, now dominated by the state $|\psi_{1}\rangle$. This smooth transition illustrates how the GP interpolates between the contributions of the two pure states as controlled by the classical weight $\lambda$.

\section{Analysis of results}
\par
In this section, we analyze the results of the GP acquired by the introduced two-mode mixed SCSs from the previous section during unitary cyclic evolution. The focus is on how the GP is influenced by various parameters such as the squeezing parameters, and the coherence parameters of the individual modes. We compare the GP of different mixed states to gain a deeper understanding of their behavior.
\par
\begin{figure}
 \centering
 \subfigure[Mixed state $\rho$ with $r_{0}=r_{1}=0$]{\includegraphics[width=5cm]{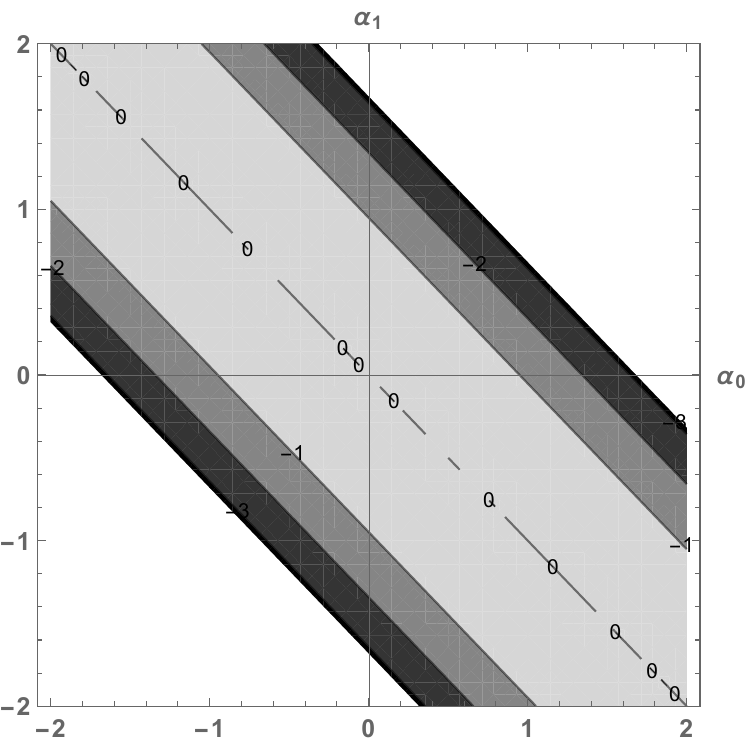}\label{4a}}
  \hfill
 \subfigure[Mixed state $\rho^{\prime}$ with $r_{0}=r_{1}=0$]{\includegraphics[width=5cm]{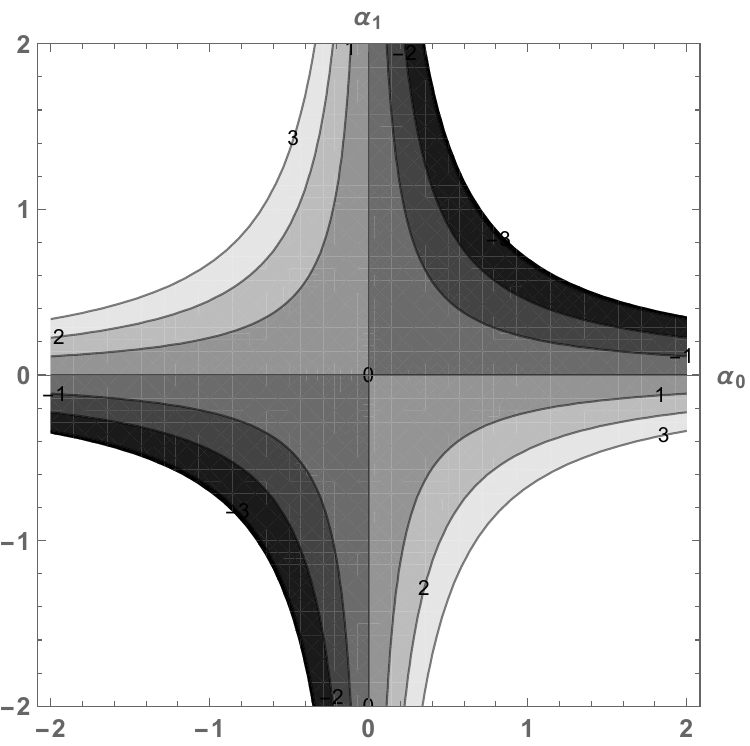}\label{4b}}
  \hfill
 \subfigure[Mixed state $\rho^{\prime \prime}$ with $r_{0}=r_{1}=0$]{\includegraphics[width=5cm]{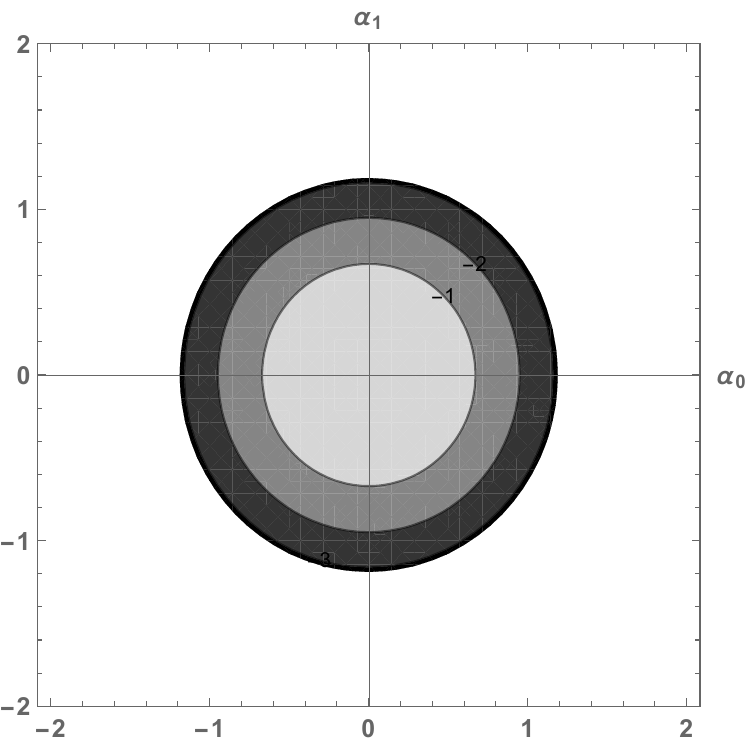}\label{4c}}
  \hfill
 \subfigure[Mixed state $\rho$ with $r_{0}=r_{1}=0.5$]{\includegraphics[width=5cm]{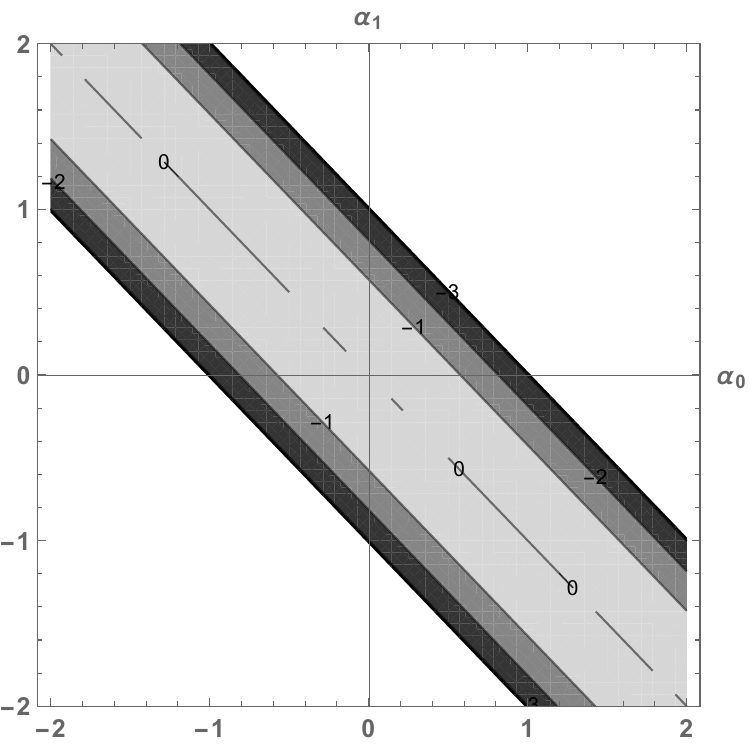}\label{4d}}
 \hfill
  \subfigure[Mixed state $\rho^{\prime}$ with $r_{0}=r_{1}=0.5$]{\includegraphics[width=5cm]{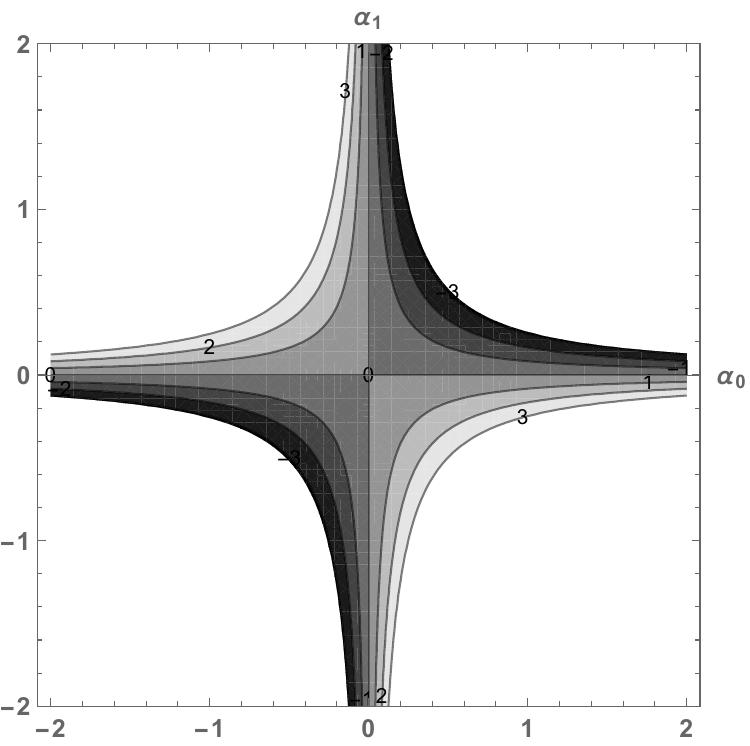}\label{4e}}
 \hfill
  \subfigure[Mixed state $\rho^{\prime \prime}$ with $r_{0}=r_{1}=0.5$]{\includegraphics[width=5cm]{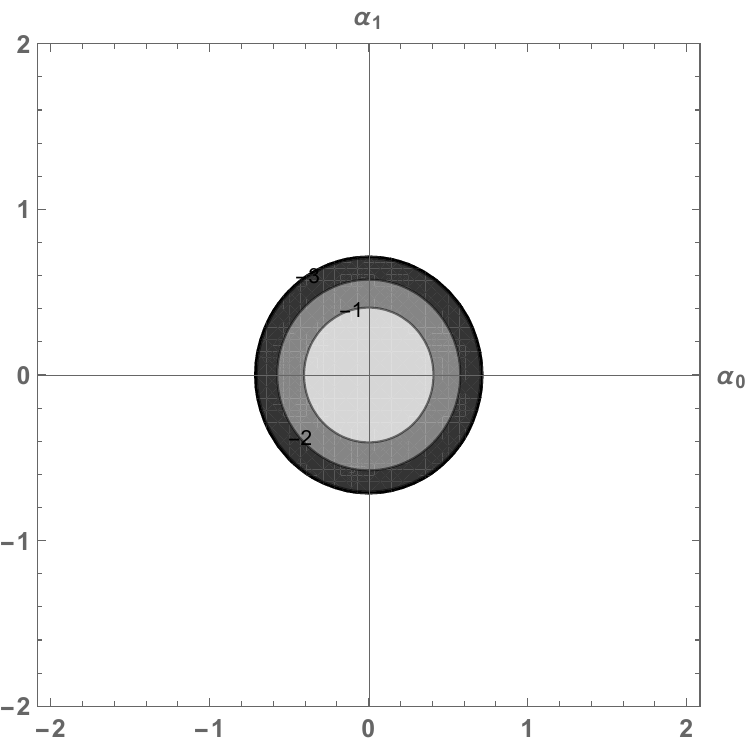}\label{4f}}
 \hfill
  \subfigure[Mixed state $\rho$ with $r_{0}=r_{1}=1$]{\includegraphics[width=5cm]{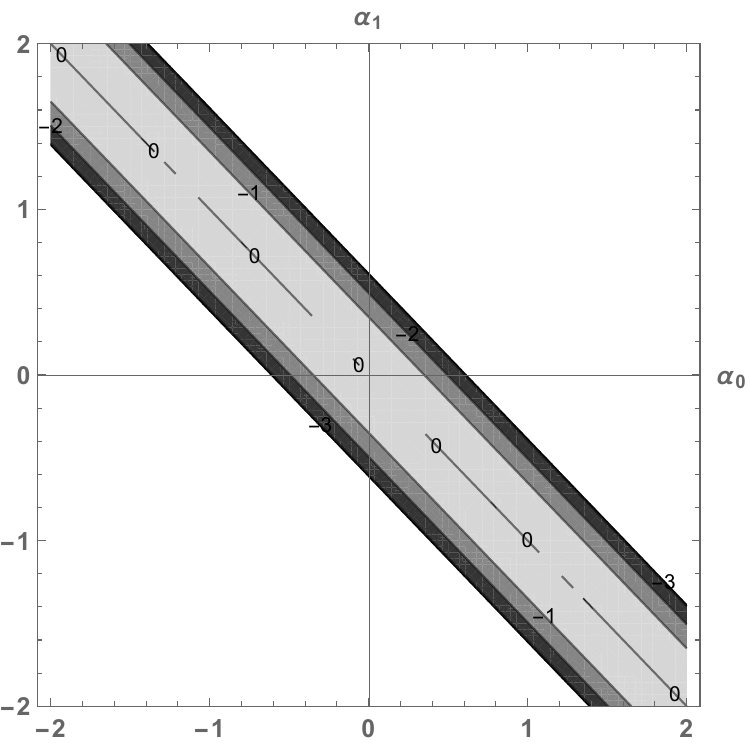}\label{4g}}
 \hfill
  \subfigure[Mixed state $\rho^{\prime}$ with $r_{0}=r_{1}=1$]{\includegraphics[width=5cm]{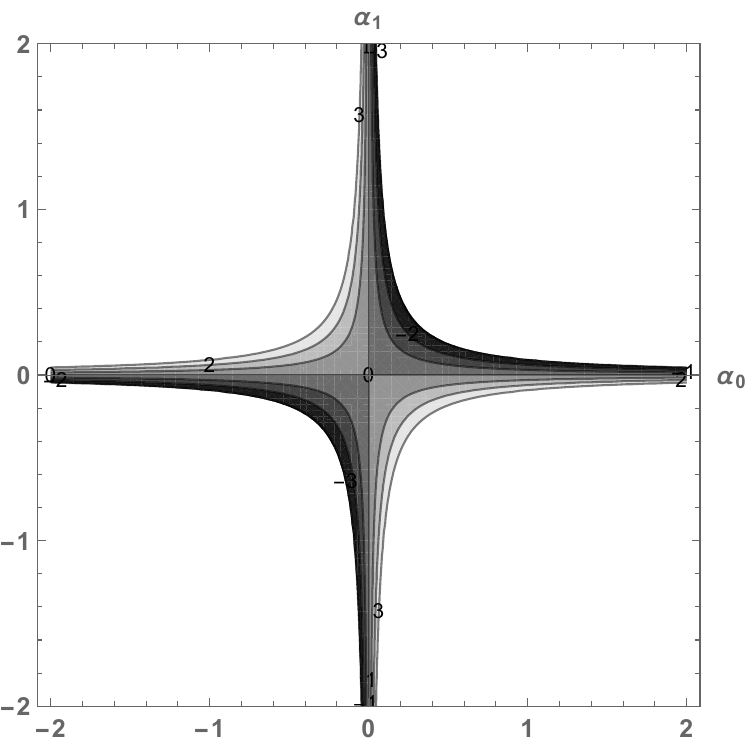}\label{4h}}
 \hfill
  \subfigure[Mixed state $\rho^{\prime \prime}$ with $r_{0}=r_{1}=1$]{\includegraphics[width=5cm]{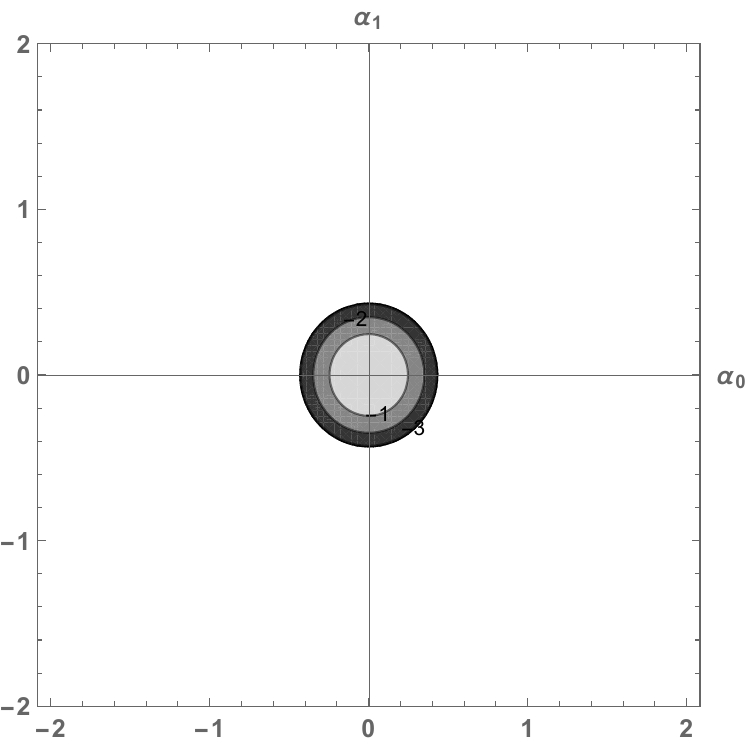}\label{4i}}
  \caption{\small{Contour plots of the GP as functions of $\alpha_{0}$ and $\alpha_{1}$, for $\theta=\pi/4$, $\lambda=1/2$ and various values of the squeezing parameters.}}\label{ContourgpSCSs}
 \end{figure}

To further explore the behavior of the GP, we compare the GP of the three mixed states introduced in the previous section. These states are characterized by different configurations of the SCSs $|\psi_{1}\rangle$ and $|\psi_{2}\rangle$. Fig. \ref{ContourgpSCSs} presents contour plots of the GP for the states $\rho$, $\rho^{\prime}$ and $\rho^{\prime \prime}$ for different values of the squeezing parameter, with respect to $\alpha_{0}$ and $\alpha_{1}$, while keeping $\theta$ constant at $\pi/4$ and $\lambda=1/2$. It can be observed that as the squeezing parameters $r_{0}$ and $r_{1}$ increase, the GP of the three states undergoes noticeable compression. The figure illustrates that by increasing the squeezing parameters, the plots for state $\rho$ are compressed linearly, for state $\rho^{\prime}$ are compressed hyperbolically and for state $\rho^{\prime \prime}$ are compressed elliptically. This analysis aligns with the general understanding that coherent states minimize uncertainty, while squeezed states enhance precision by adjusting uncertainty \cite{gerry2005introductory}. The results demonstrate that the impact of this uncertainty adjustment in squeezed states is clearly visible when studying the GP in these states. The manipulation and control of the GP in mixed states can enhance quantum measurement precision, benefiting applications in quantum sensing and metrology.

\par
As seen in the equations (\ref{gp-mixstate1}, \ref{gp-mixstate2}, and \ref{gp-mixstate3}), the GP behaves differently for separable and entangled mixed SCSs. In the case of separable mixed SCSs, such as $\rho^{\prime}$ and $\rho^{\prime \prime}$, the GP demonstrates a relatively straightforward, exponential dependence on the squeezing parameters. This indicates a simpler relationship where the phase evolves predictably as the squeezing parameters vary. However, the situation is notably more complex in the entangled mixed SCS $\rho$. In this state, the GP is not solely determined by the squeezing parameters. Instead, it is influenced by more intricate interactions between the states involved in the entanglement, represented by $p_{01}$.
\par
\begin{figure}
 \centering
 \subfigure[$\lambda=0$]{\includegraphics[width=5cm]{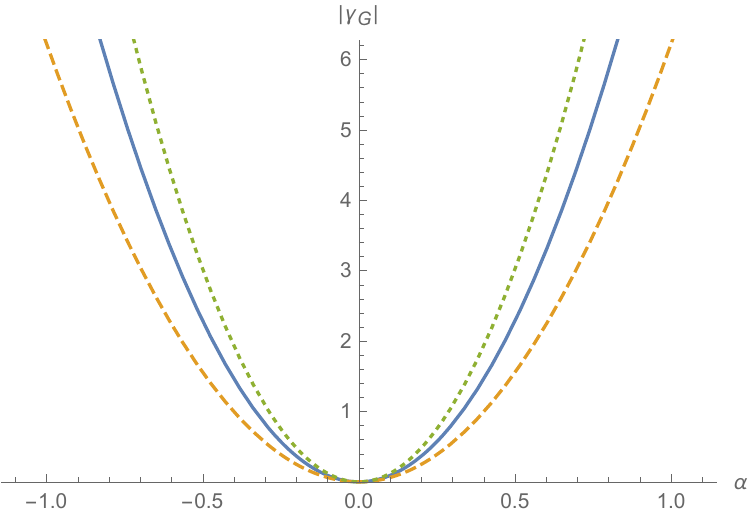}\label{5a}}
  \hfill
 \subfigure[$\lambda=1/4$]{\includegraphics[width=5cm]{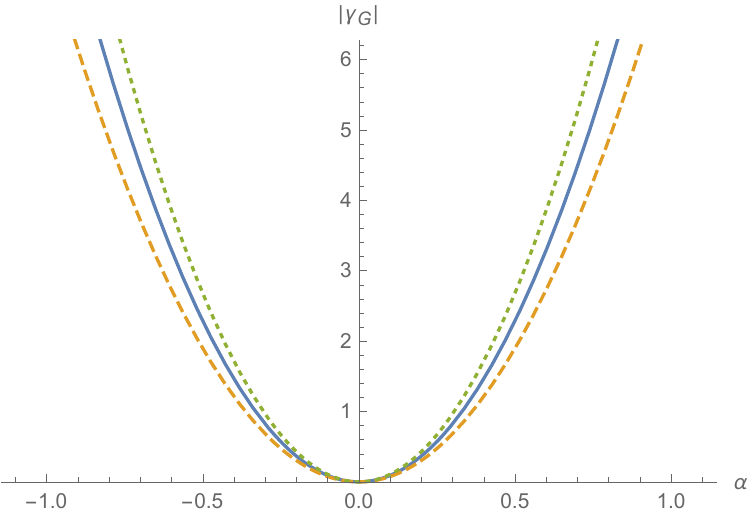}\label{5b}}
  \hfill
 \subfigure[$\lambda=1/2$]{\includegraphics[width=5cm]{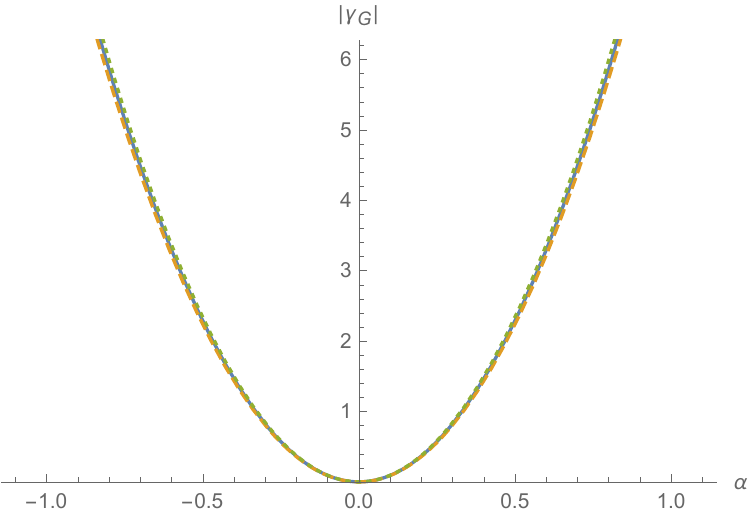}\label{5c}}
  \hfill
 \subfigure[$\lambda=3/4$]{\includegraphics[width=5cm]{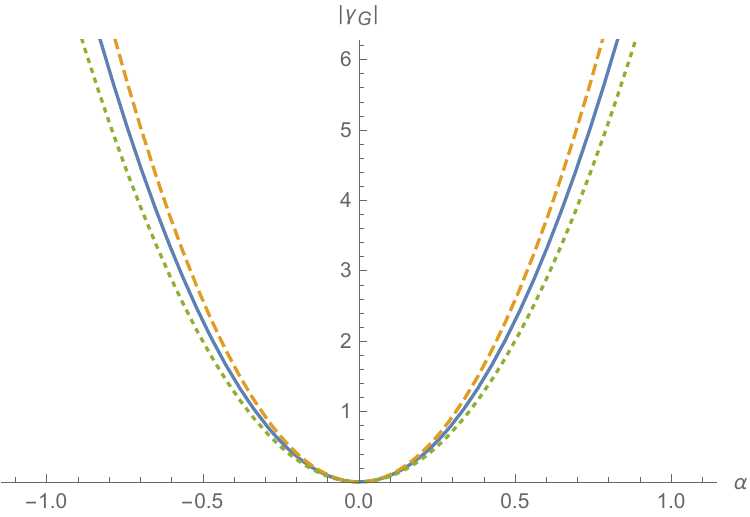}\label{5d}}
\hspace{1.5 em}
  \subfigure[$\lambda=1$]{\includegraphics[width=5cm]{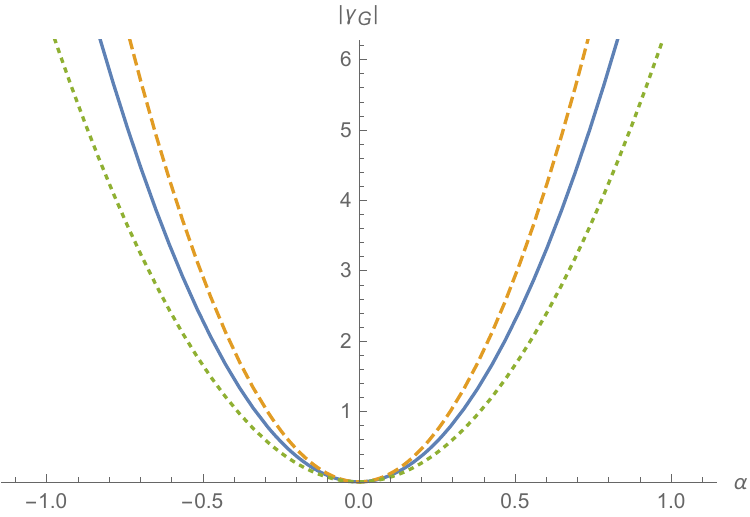}\label{5e}}
\caption{\small{The modulus of the of the GPs for $\rho$ (solid line), $\rho^{\prime}$ (dashed line), and $\rho^{\prime \prime}$ (dotted line) is plotted as a function of $\alpha_{0} = \alpha_{1} = \alpha$, with $\theta = \pi/4$, $r_{0} = 0.2$, and $r_{1} = 0.5$, for different values of the classical weight.}}\label{ComparingGPs}

 \end{figure}

The effect of the classical weight on the GP for these states is illustrated in Figures \ref{ContourgpSCS1}, \ref{ContourgpSCS2}, and \ref{ContourgpSCS3}. In the contour plots, as the parameter $\lambda$ changes, the GP for the state $\rho^{\prime}$ consistently retains an elliptic-like form, while for the state $\rho^{\prime \prime}$, it maintains a hyperbolic-like form. This stability in shape for $\rho^{\prime}$ and $\rho^{\prime \prime}$ indicates that the GP in these states is less sensitive to variations in $\lambda$. In contrast, for the entangled mixed state $\rho$, the shape of the GP varies significantly with changes in $\lambda$, as clearly illustrated in the figures. This variability highlights the strong dependence of the GP on $\lambda$ in entangled mixed state, where small changes can lead to noticeable shifts in the GP. To extend our investigation, Fig.\ref{ComparingGPs} compares the GP of separable and entangled mixed SCSs, with the modulus of the GP plotted as a function of $\alpha$. In this comparison, we assume $\alpha_0 = \alpha_1 = \alpha$ and fix the parameters $\theta = \pi/4$, $r_{0 }= 0.2$, and $r_{1 }= 0.5$. The GP is analyzed for different values of the classical weight. The solid line in the figures represents the GP of $\rho$, the dashed line corresponds to the GP of $\rho^{\prime}$, and the dotted line represents the GP of $\rho^{\prime \prime}$. It is evident that there is a symmetry line at $\alpha=0$, and the GP increases as the absolute value of $\alpha$ grows. The behavior of the GP with respect to the classical weight $\lambda$ shows different trends depending on the state. For the entangled state $\rho$, the GP remains unaffected by variations in $\lambda$. However, for the separable states $\rho^{\prime}$ and $\rho^{\prime\prime}$, the GP shows a distinct dependency on $\lambda$. Specifically, as $\lambda$ increases for a fixed $\alpha$, the GP of $\rho^{\prime}$ increases, while the GP of $\rho^{\prime\prime}$ decreases.

\section{Summary and conclusion}
\par
In summary, we investigated the GP acquired by two-mode mixed SCSs during unitary cyclic evolution. We examined different configurations of these mixed states and analyzed how various parameters, such as squeezing and coherence parameters, affect the GP.
Our analysis revealed that the classical weight parameter $\lambda$ controls the interpolation between different configurations of the SCSs, affecting the GP's behavior.
We have examined the influence of squeezing parameters on the GP, revealing dependencies and unique patterns of compression in the GP contours for different states. The GP of the mixed states $\rho$, $\rho^{\prime}$, and $\rho^{\prime \prime}$ exhibited distinct compression patterns—linear, hyperbolic, and elliptical, respectively—as the squeezing parameters $r_{0}$ and $r_{1}$ increased. The results aligned with the understanding that coherent states minimize uncertainty, while squeezed states enhance precision by adjusting uncertainty. This adjustment was clearly visible in the GP behavior of the mixed states. The sensitivity of the GP to squeezing parameters can be leveraged in applications requiring precise control of the GP, such as quantum information processing, quantum metrology, and quantum interferometry.

\printbibliography 

\end{document}